%% file: main.tex
\documentclass[cameraready]{Interspeech}

\title{RT-Tango: Real-Time Distributed Binaural Speech Enhancement for Low-Power Hearing Aid Devices}

\author[affiliation={1,2}, orcid=0009-0005-9846-9767]{Zahra}{Benslimane}
\author[affiliation={1}]{Pierre}{Chouteau}
\author[affiliation={1}, orcid=0000-0002-5102-7735]{Martyna}{Poreba}
\author[affiliation={1}, orcid= 0000-0002-4441-9239]{Fabrice}{Auzanneau}
\author[affiliation={1}, orcid=0009-0000-9061-4396]{Michal}{Szczepanski}
\author[affiliation={1}, orcid=0000-0003-1257-2607]{Fabian}{Chersi}
\author[affiliation={2}, orcid=0000-0002-6848-0114]{Romain}{Serizel}

\address{
    $^1$ Université Paris-Saclay, CEA, List, F-91120 Palaiseau, France \\
    $^2$ Université de Lorraine, CNRS, Inria, LORIA, F-54000 Nancy, France
}

\email{zahra-hafida.benslimane@cea.fr}


\usepackage{comment}

\newcommand{\todo}[1]{%
  \textcolor{red}{[TODO: #1]}%
}

\newcommand{\zahra}[1]{\textcolor{blue}{\small {[Zahra : #1]}}}

\usepackage{placeins}  

\makeatletter
\def\blfootnote{\gdef\@thefnmark{}\@footnotetext}
\makeatother

\begin{document}


\maketitle

\blfootnote{This research was carried out with the support of the French National Research Agency as part of the REFINED project, “REal-time artiFicial INtelligence for hEaring aiDs” (ANR21-CE19-0043).}

\begin{abstract}
    Real-time binaural speech enhancement is constrained by latency, computational cost, and inter-device communication, yet existing efficient solutions predominantly address single-channel settings. In this paper, we introduce RT-Tango, a real-time distributed binaural speech enhancement framework designed for streaming on resource-constrained platforms and specifically for hearing aids. RT-Tango relies on a two-stage distributed architecture combining perceptually motivated ERB feature compression, lightweight grouped recurrent mask estimation, and temporal sparsification to reduce computational cost. Stringent latency constraints are addressed by decoupling spectral resolution from algorithmic delay using an asymmetric STFT, together with causal recurrent inference and online estimation of spatial statistics. Experimental results show that RT-Tango achieves competitive speech enhancement while significantly reducing MACs operations and functioning at ultra-low latencies as low as 8 ms.
    
\end{abstract}

\input{1introduction}

\input{2method}

\input{3experiments}
\input{5conclusions}

\input{main.bbl}
\end{document}

%% file: 1introduction.tex
\section{Introduction}


Real-time speech enhancement (SE) in bandwidth-limited binaural and distributed systems, such as hearing aids, must operate under strict latency and computational constraints. Each device processes audio locally, exchanging only minimal information, which places strong limitations on model complexity and communication. Addressing SE under these joint constraints is essential for practical on-device deployment.

One line of research focuses on reducing the computational cost of on-device SE through neural model compression, typically combining structured pruning and quantization to reduce memory use and computational complexity \cite{9437977,8892545,fedorov_tinylstm_2020}. Hardware-oriented studies further show that not only the model's sparsity level but also its structure critically impact the throughput-memory-quality trade-off on embedded platforms \cite{stamenovic2021weightblockunitexploring}. However, SE models are particularly sensitive to aggressive quantization due to their regression-based objectives, which amplifies quantization noise and hinders full integer deployment \cite{cohen23_interspeech}. This has motivated the search for alternative solutions, such as hardware-aware mixed-precision inference \cite{rusci2022acceleratingrnnbasedspeechenhancement} or distillation-based training strategies \cite{10591369}. Complementary research explores architectural efficiency: designing compact models that inherently require fewer operations. Grouped processing is attractive for SE, where feature maps are partitioned into small groups and processed in parallel by tiny sub-networks that periodically exchange information \cite{8910352, 9414322}. Other works reduce input dimensionality by sub-sampling \cite{10095700} or using an Equivalent Rectangular Bandwidth (ERB)-scaled filterbank to compress the frequency representations, as in GTCRN \cite{rong_gtcrn_2024}.

Despite these advances, most efficiency-driven approaches focus on single-microphone SE. In contrast, efficient multi-microphone methods remain comparatively underexplored, even though they can effectively exploit spatial cues for improved noise reduction.
Recent works have begun to address this gap by proposing computationally efficient multi-microphone architectures, either by decoupling spatial and spectral processing \cite{pandey_tgru_2025} or by employing lightweight attention mechanisms to model spatial dependencies \cite{10889867}. Nevertheless, these approaches typically incur higher computational cost than ultra-lightweight monaural models and often assume a centralized processing with access to all microphone signals. In practical binaural hearing-device systems, microphones are distributed across physically separated nodes, and centralized processing would require continuous high-bandwidth wireless transmission, which is infeasible under severe energy and bandwidth constraints. To address this limitation, fully distributed fusion-center-free signal processing frameworks have been explored to balance SE performance with bandwidth-efficient inter-device communication \cite{5483092,4740158}.

While prior work has demonstrated the feasibility of reduced-bandwidth distributed binaural SE \cite{9466439}, jointly achieving low latency and low computational complexity remains an open challenge. To the best of our knowledge, no existing approach combines the required design elements within a distributed binaural framework under explicit real-time latency and computational constraints. This paper addresses this gap by introducing RT-Tango, a binaural SE framework specifically designed for real-time deployment on resource-constrained hearing devices. Concretely, our framework relies on the following efficiency-driven design choices:

\begin{itemize}
    \item Perceptually motivated feature compression using an ERB-scaled filterbank, reducing input dimensionality while preserving speech-relevant cues.
    \item Lightweight grouped recurrent mask estimation, enabling parallel and partitioned processing to reduce computational cost in distributed binaural settings.
    \item Decoupled frequency resolution and algorithmic latency reduction through an asymmetric STFT configuration tailored for streaming operations.
    \item Temporal sparsification for real-time inference, including both learned skip-RNN gating and fixed-rate frame reuse to reduce recurrent updates.
\end{itemize}

%% file: 2method.tex
\section{Methodology}
\label{sec:method}
RT-Tango revisits the Tango framework~\cite{9466439} from a system-level perspective. This redesign is motivated by the stringent constraints of hearing aids, where strict causality, ultra-low latency, limited memory, and low-power embedded hardware are critical requirements. Rather than introducing isolated optimizations, RT-Tango adopts complementary architectural and algorithmic design principles detailed in the following sections.

\subsection{Baseline}
\begin{figure}[t]
  \centering
  \includegraphics[width=\linewidth]{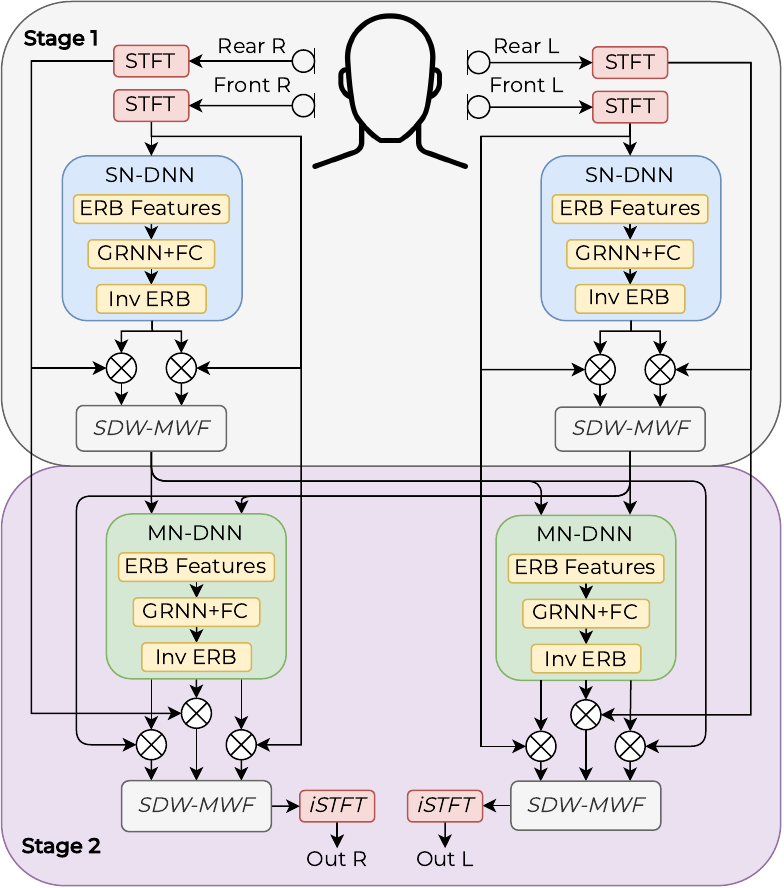}
  \caption{ Block diagram of RT-Tango. It preserves Tango's two-stage processing scheme while introducing additional or modified modules focused on efficiency (highlighted in yellow) and low-latency streaming (in italics).}
  \label{fig:tango_diag}
\end{figure}

Tango serves as our baseline due to its two-stage distributed architecture (illustrated in Figure~\ref{fig:tango_diag}) and proven effectiveness in spatially unconstrained microphone configurations~\cite{9466439}. Each ear-node independently estimates speech and noise masks using a Single-Node DNN (SN-DNN) and computes a Speech Distortion Weighted Multichannel Wiener Filter (SDW-MWF)~\cite{10.1016/j.specom.2007.02.001}. This produces an ear-specific compressed signal, which is transmitted to the contra-lateral ear-node. Then, a Multi-Node DNN (MN-DNN) refines the masks using both local signals and exchanged representations, from which a final SDW-MWF generates the enhanced binaural output. In its original formulation, however, the baseline imposes no latency or complexity constraints.


\subsection{Lightweight neural mask estimation}
To reduce the computational cost of this distributed binaural processing, RT-Tango improves the efficiency in neural mask estimation by means of perceptually driven feature compression, grouped modeling of localized spectral dependencies, and temporal sparsification, while preserving the characteristic two-stage distributed spatial filtering framework of Tango.

\subsubsection{Feature compression}
The ERB representation is used as the front-end for both neural mask estimation networks (indicated as SN-DNN and MN-DNN in Figure~\ref{fig:tango_diag}). This perceptually motivated transformation reduces spectral resolution at high frequencies, where human auditory sensitivity is coarser, while preserving finer resolution at lower frequencies that carry most speech energy and cues. By aligning the features with perceptual frequency scaling, ERB compression reduces the input dimensionality of the mask estimators, lowering the number of parameters and operations in subsequent layers. The estimated masks are then mapped back to the linear-frequency domain using an inverse ERB transformation to restore the spectral resolution required for filtering.

\subsubsection{Grouped RNN mask estimation}
The mask estimation networks of RT-Tango implement a grouped recurrent neural network (GRNN) to reduce computational complexity while preserving modeling capacity~\cite{gao-etal-2018-efficient}. This choice is motivated by the localized nature of spectral dependencies in speech signals and the need to control recurrent computational complexity. Localized recurrent modeling within frequency sub-bands is consistent with common SE assumptions as spectral dependencies are typically stronger within nearby frequency regions. Since the computational cost of recurrent layers scales approximately quadratically as $\mathcal{O}(H^2)$ with the hidden state dimension $H$, partitioning the feature space into $G$ groups of size $H/G$, the total recurrent complexity becomes $G \cdot \mathcal{O}((H/G)^2) = \mathcal{O}(H^2/G)$, making it significantly lower compared to a full-band RNN. Although the processing is performed independently within each group, cross-band dependencies are captured through a representation rearrangement mechanism \cite{gao-etal-2018-efficient}, enabling cross-band information exchange and global spectral modeling. 
\input{table1}

\subsubsection{Temporal sparsification}
To further reduce the computational load of high-rate real-time streaming, RT-Tango exploits temporal redundancy in the input signal using skip inference strategies. Instead of performing a neural inference at every frame, the mask estimation networks reuse previously estimated masks when the input signal exhibits limited temporal variation. This temporal sparsification reduces the mask's update frequency and thus the overall computational cost during streaming inference. In RT-Tango, we define Fixed-Rate Skipping (FRS), as a strategy that executes the mask estimator at a predefined interval and reuses previously estimated masks in between. We compare this with learned skip gates \cite{fedorov_tinylstm_2020,campos2018skip}, where a gating mechanism dynamically decides whether to update the recurrent state at each frame (see Section~\ref{sec:ablation_skip}).

\subsection{Low-latency streaming}
To support low-latency streaming, we use an asymmetric STFT configuration~\cite{9961936, Wood_2019}. We use a long analysis window (aW) that preserves the frequency resolution and a shorter synthesis window (sW) to reduce the reconstruction latency, enabling low-latency real-time streaming. The spatial covariance matrices (SCM) required by the SDW-MWF are updated online using a recursive exponential moving average (EMA). Under these streaming constraints, the system operates as RT-Tango-OS.

%% file: table1.tex

\begin{table*}[t]
        
        \caption{Model comparison with left/right ear scores. Computational cost is reported in MMACs/s and decomposed into DNN and SDW-MWF components. A hop size of $\SI{16}{\milli\second}$ corresponds to $\approx 62$ STFT frames/s, while $\SI{4}{\milli\second}$ corresponds to $\approx 250$ STFT frames/s. SI-SAR is not reported for the unprocessed mixture, where no processing artifacts are present.}
        \label{tab:lr_metrics}
        \centering
        \fontsize{8.5}{9.5}\selectfont
        \setlength{\tabcolsep}{4.9pt}
        
        \begin{tabular}{lcccccccccccccc}
        \toprule
        & \multicolumn{1}{c}{\textbf{STFT}} 
        & \multicolumn{3}{c}{\textbf{MMACs/s} $^\downarrow$} 
        & \multicolumn{2}{c}{\textbf{SI-SIR} $^\uparrow$} 
        & \multicolumn{2}{c}{\textbf{SI-SDR} $^\uparrow$} 
        & \multicolumn{2}{c}{\textbf{SI-SAR} $^\uparrow$} 
        & \multicolumn{2}{c}{\textbf{STOI} $^\uparrow$} 
        & \multicolumn{2}{c}{\textbf{PESQ} $^\uparrow$} \\
        \cmidrule(lr){3-5}
        \cmidrule(lr){6-7}\cmidrule(lr){8-9}\cmidrule(lr){10-11}
        \cmidrule(lr){12-13}\cmidrule(lr){14-15}
        
        \textbf{Model} &
        \textbf{Hop} &
        \textbf{Total} & \textbf{DNN} & \textbf{SDW-MWF} &
        \textbf{L} & \textbf{R} &
        \textbf{L} & \textbf{R} &
        \textbf{L} & \textbf{R} &
        \textbf{L} & \textbf{R} &
        \textbf{L} & \textbf{R} \\
        \midrule
        
        Unprocessed
        & $-$
        & $-$ & $-$ & $-$
        & $0.0$ & $- 4.0$
        & $- 0.6$ & $- 4.6$
        & $-$ & $-$
        & $0.68$ & $0.56$
        & $1.14$ & $1.10$ \\
        \midrule
        
        Tango
        & $\SI{16}{\milli\second}$
        & $605.98$ & $604.5$ & $1.48$
        & $20.8$ & $24.1$
        & $4.2$ & $4.4$
        & $4.7$ & $4.7$
        & $0.83$ & $0.84$
        & $1.61$ & $1.64$ \\
        
        GTCRN
        & $\SI{16}{\milli\second}$
        & $48.98$ & $48.98$ & $-$
        & $16.1$ & $14.1$
        & $5.6$ & $3.7$
        & $6.4$ & $4.5$
        & $0.76$ & $0.69$
        & $1.47$ & $1.34$ \\
        
        GTCRN
        & $\SI{4}{\milli\second}$
        & $197.5$ & $197.5$ & $-$
        & $16.6$ & $13.8$
        & $6.0$ & $4.0$
        & $6.7$ & $5.0$
        & $0.79$ & $0.71$
        & $1.52$ & $1.36$ \\
        \midrule
        
        Tango-RNN
        & $\SI{16}{\milli\second}$
        & $67.20$ & $65.72$ & $1.48$
        & $21.6$ & $25.0$
        & $4.7$ & $5.0$
        & $5.2$ & $5.2$
        & $0.84$ & $0.85$
        & $1.66$ & $1.70$\\\noalign{\vskip 0.1em} 

        \hspace{0.5em} + \textnormal{GRNN}$_{\scriptscriptstyle \mathrm{MN}=2}^{\scriptscriptstyle \mathrm{SN}=8}\ \star$
        & $\SI{16}{\milli\second}$
        & $18.22$ & $16.74$ & $1.48$
        & $21.3$ & $24.8$
        & $4.5$ & $4.8$
        & $5.1$ & $5.0$
        & $0.84$ & $0.84$
        & $1.66$ & $1.70$ \\
        \midrule
        
        \textbf{RT-Tango} (ours)
        & $\SI{4}{\milli\second}$
        & $\mathbf{33.41}$ & $28.08$ & $5.33$ &
        $20.8$ & $24.6$ &
        $4.4$ & $4.7$ &
        $5.0$ & $5.0$ &
        $0.84$ & $0.84$ &
        $1.66$ & $1.71$ \\

        \textbf{RT-Tango-OS} (ours)
        & $\SI{4}{\milli\second}$
        & $\mathbf{35.14}$ & $28.08$ & $7.01$
        & $20.5$ & $24.7$
        & $2.9$ & $3.8$
        & $3.4$ & $4.0$
        & $0.80$ & $0.82$
        & $1.54$ & $1.63$ \\
    
    \bottomrule
    \end{tabular}
    
\end{table*}

%% file: 3experiments.tex
\section{Experiments}
\label{sec:experiment}

\subsection{Experimental setup}

Models were trained on a simulated binaural dataset following the protocol of Monir et al.~\cite{monir2025frequencyweightedtraininglossesphonemelevel}, which utilized a four-microphone hearing aid configuration with two microphones per ear. Clean speech from LibriSpeech~\cite{7178964} was mixed with speech-shaped noise and real-world environmental noise. Evaluation was conducted on a subset of the BinauRec\footnote{Online: https://zenodo.org/records/7256984} binaural dataset~\cite{delebecque_binaurec_2023}, comprising 1,200 mixtures generated using measured room impulse responses (RIRs) acquired using a portable hearing laboratory (PHL)~\cite{pavlovic_high-fidelity_2019} with behind-the-ear hearing aids mounted on a dummy head. We considered configurations where the target speech source was located in front and the noise positioned at $45^\circ$ and $90^\circ$ to the right of the target, respectively. This results in a lower SIR at the right ear than at the left ear. Dry signals are mixed at input SNRs of $-5$, $0$, and $5$~dB, before being convolved with the RIR. 

Enhancement performance was evaluated using scale-invariant objective metrics (SI-SDR, SI-SIR, and SI-SAR), reported in dB~\cite{8683855}, together with perceptual measures PESQ~\cite{941023} and STOI~\cite{5495701}. Processing cost is reported in terms of MACs either per second or per frame, measured for one processing node. Reported MACs include the DNN mask estimators, ERB and inverse-ERB transformations, and SDW-MWF processing, while FFT and iFFT operations are excluded.

Alongside the original CNN-based Tango implementation, we  consider a causal variant, Tango-RNN, in which both the SN-DNN and MN-DNN are implemented using two stacked stateful RNN layers with 128 hidden units, followed by a fully connected layer. We also include a lightweight neural baseline by adapting GTCRN~\cite{rong_gtcrn_2024} to the distributed setting, deploying one instance per node. Each instance processes the four microphone signals as input and applies its estimated mask to the local reference channel, preserving a distributed per-node architecture. 

Models were implemented in PyTorch and trained with an Adam optimizer (lr = $10^{-3}$), by minimizing the mean squared error between the estimated time–frequency mask and the target ideal ratio mask. Training audio used a 32~ms STFT analysis window with a 16~ms hop for grouped ablations, FRS, and RT-Tango/RT-Tango-OS. For learned skipping and GTCRN (4~ms), the hop was reduced to 4~ms to match RT-Tango's frame rate.

RT-Tango replaces the recurrent layers in Tango-RNN with GRNNs to reduce its computational cost. We use $G=8$ groups for the SN-DNN and $G=2$ groups for the MN-DNN. Temporal sparsification is implemented using FRS, with update rates of $1/4$ for the SN-DNN and $1/2$ for the MN-DNN, corresponding to one inference every four and two frames, respectively.

All results, except those obtained for RT-Tango-OS, were computed using SCMs estimated over entire utterances. RT-Tango-OS instead uses recursive SCM updates with a forgetting factor $\alpha = 0.995$, applied every 8 frames at the 4 ms input rate, resulting in an effective update interval of 32~ms ($\approx$ 31 updates/s). As online SCM estimation requires an adaptation period, RT-Tango-OS was evaluated in steady state by repeating each mixture a sufficient number of times to allow for convergence. The metrics are computed on the final repetition (see Section~\ref{sec:online_psd_results}).

\input{table2}
\input{4results}



%% file: table2.tex
\begin{table}[!ht]
  \caption{Ablation study on grouping strategies. The reported MMACs per frame are computed exclusively for the DNNs.} 
  \label{tab:ablation_groups}
  \centering
  \fontsize{8.5}{9.5}\selectfont
  \setlength{\tabcolsep}{4pt}

  \begin{tabular}{lccccc}
    \toprule

    \multicolumn{1}{c}{} &
    \multicolumn{1}{c}{\textbf{DNN (SN+MN)}} &
    \multicolumn{2}{c}{\textbf{SI-SDR}} &
    \multicolumn{2}{c}{\textbf{SI-SAR}} \\
    \cmidrule(lr){3-4}\cmidrule(lr){5-6}

    \textbf{Model} &
    \textbf{MMACs / frame} &
    \textbf{L} & \textbf{R} &
    \textbf{L} & \textbf{R} \\
    \midrule
    
    Tango-RNN & $1.06$ &
    $4.7$ & $5.0$ &
    $5.2$ & $5.2$ \\
    
    \midrule

    \hspace{0.5em}\textnormal{+ GRNN}$_{\scriptstyle \textnormal{SN}=2}$ & $0.69$ &
    $4.7$ & $4.9$ &
    $5.2$ & $5.2$ \\
    
    \hspace{0.5em}\textnormal{+ GRNN}$_{\scriptstyle \textnormal{SN}=8}$ & $0.59$ &
    $4.7$ & $4.9$ &
    $5.1$ & $5.1$ \\
    
    \midrule

    \hspace{0.5em}\textnormal{+ GRNN}$_{\scriptstyle \textnormal{MN}=2}$ & $0.74$ &
    $4.7$ & $4.9$ &
    $5.2$ & $5.1$ \\
    
    \hspace{0.5em}\textnormal{+ GRNN}$_{\scriptstyle \textnormal{MN}=8}$ & $0.64$ &
    $3.8$ & $4.1$ &
    $4.2$ & $4.3$ \\
    \bottomrule
  \end{tabular}
\end{table}

%% file: 4results.tex
\subsection{Results}
\label{sec:results}


In Table~\ref{tab:lr_metrics}, we report the results of RT-Tango and its streaming variant RT-Tango-OS, and compare them to the Tango baseline, Tango-RNN, and GTCRN. Compared to GTCRN, both Tango and Tango-RNN achieve consistently higher PESQ, STOI, and SI-SIR scores across both ears. Although GTCRN slightly outperforms them in SI-SDR and SI-SAR on the less noise-affected left node, Tango-based approaches show a more balanced left-right behavior due to their two-stage mechanism. This interaural balance is particularly desirable in hearing-aid applications for stable spatial perception and listening comfort.

Tango-RNN reduces Tango’s complexity to 67.2 MMAC/s, while RT-Tango reaches 33.4~MMAC/s despite operating at a higher frame rate. Under identical conditions, RT-Tango is nearly six times more efficient than GTCRN (197.5 MMAC/s). This robustness to compression stems from its hybrid architecture, where neural networks guide the spatial filter rather than directly reconstructing the signal.
 
RT-Tango-OS enables strictly causal streaming with an algorithmic latency of 8~ms, through the asymmetric STFT and online SDW-MWF updates, resulting in a slight but expected performance degradation relative to offline RT-Tango. RT-Tango-OS requires 0.49 MMACs per frame (0.27 MMACs for the DNN and 0.22 MMACs for the SDW-MWF update). With the proposed temporal sparsification, it corresponds to 35.14 MMACs/s at $\approx$250 frames/s. Without sparsification (not reported here), the computational cost would reach 124 MMACs/s. The RT-Tango-OS configuration results from the progressive design choices described in Section~\ref{sec:method}, whose individual contributions are analyzed in the ablation studies below.

\input{table3}
\input{table4}

\subsubsection{Impact of Grouped Recurrent Modeling}
Grouping applied independently to SN-DNN and MN-DNN reveals different behaviors (Table~\ref{tab:ablation_groups}). Grouping the SN-DNN significantly reduces cost with negligible impact on quality. Using 8 groups on SN-DNN lowers the total DNN (SN+MN) complexity from 1.06 to 0.59 MMAC/frame while maintaining SI-SDR at 4.7/4.9 dB and preserving PESQ/STOI. In contrast, MN-DNN is more sensitive to grouping. While MN=2-GRNN preserves performance, MN=8-GRNN degrades SI-SDR and SI-SAR by around 0.8-1 dB. These results justify the combined grouping strategy (SN=8, MN=2), marked with ($\star$) in Table~\ref{tab:lr_metrics}.

\subsubsection{Effect of temporal sparsification via skip inference}
\label{sec:ablation_skip}

Tables~\ref{tab:skip_single} and~\ref{tab:skip_multi} evaluate temporal sparsification for the SN-DNN and MN-DNN stages. For SN-DNN, FRS preserves performance within 0.2 dB of the baseline for both 1/2 and 1/4 configurations while eliminating computation on non-updated frames. Learned skipping relies on a gating mechanism that must be evaluated at every frame, introducing additional MACs. Learned skipping achieves an effective skip ratio of 80\%, but still introduces minor degradation, primarily on the right channel. The MN-DNN is more sensitive to learned skipping; it reduces SI-SDR from 4.5 dB to 3.8 dB (SkipRNN \cite{campos2018skip}) and 3.3 dB (TinyLSTM \cite{fedorov_tinylstm_2020}) on the left channel, whereas FRS maintains performance within 0.2 dB of the baseline. Given its stability and predictable computational savings, FRS is adopted in the RT-Tango configuration (SN 1/4, MN 1/2), corresponding to 75\% frame skipping in the SN-DNN and 50\% in the MN-DNN. At a 4 ms hop rate, this reduces the DNN cost from 67.5 MMACs/s without sparsification to 28.08 MMACs/s.

\subsubsection{Analysis of the latency-quality trade-off}
\label{sec:low_latency}
Table~\ref{tab:lr_metrics_GRNN} analyzes the impact of reducing the synthesis window while keeping the analysis window fixed at 32~ms. These results are computed with batch (offline) SCMs to isolate the effect of the STFT configuration from online SCM estimation. Standard square-root Hann windows degrade rapidly at short synthesis lengths. In contrast, the asymmetric Hann~\cite{Wood_2019} configuration maintains better SI-SDR and SI-SAR at low latencies. An 8~ms synthesis window offers the most favorable compromise between delay and SE quality. Reducing the synthesis window to 4~ms further decreases latency but results in performance degradation. We therefore select the 8~ms asymmetric Hann synthesis window for RT-Tango-OS.

\subsubsection{Online SCM matrix estimation and convergence}
\label{sec:online_psd_results}

As can be seen in Table~\ref{tab:lr_metrics}, RT-Tango-OS achieves an SI-SIR comparable to the offline RT-Tango. However, due to the constraints imposed by low-latency streaming, the SI-SDR and SI-SAR show a slight decrease. This performance gap is first introduced by the asymmetric STFT configuration (Section \ref{sec:low_latency}), and is then amplified by the requirements of online SCM estimation. Despite this, RT-Tango-OS maintains STOI and PESQ values close to GTCRN at the same high frame rate, while still being significantly more computationally efficient. This indicates that although strictly online SCM estimation affects reconstruction, the resulting speech intelligibility and perceived quality remain competitive with our baselines.

%% file: table3.tex
\begin{table}[t]
  \caption{Temporal sparsification in the SN-DNN of ($\star$): performance and per-frame cost for updated (w/o skip) and non-updated (w/skip) frames.}
  \label{tab:skip_single}
  \centering
  \small
  \fontsize{8.5}{9.5}\selectfont

  \setlength{\tabcolsep}{4pt}


  \begin{tabular}{lcccccc}
    \toprule

    \multicolumn{1}{c}{} &
    \multicolumn{2}{c}{\textbf{MMACs / frame}} &
    \multicolumn{2}{c}{\textbf{SI-SDR}} &
    \multicolumn{2}{c}{\textbf{SI-SAR}} \\
    \cmidrule(lr){2-3}\cmidrule(lr){4-5}\cmidrule(lr){6-7}

    \textbf{Model} &
    \textbf{w/o Skip} & \textbf{w/ Skip} &
    \textbf{L} & \textbf{R} &
    \textbf{L} & \textbf{R} \\
    \midrule

    SN-DNN of ($\star$) & $0.09$ & $-$ &
    $4.5$ & $4.8$ &
    $5.1$ & $5.0$ \\
    
    \midrule
    
    ($\star$) + FRS 1/2 & $0.09$ & $0$ &
    $4.4$ & $4.7$ &
    $5.1$ & $5.0$ \\
    
    ($\star$) + FRS 1/4 & $0.09$ & $0$ &
    $4.3$ & $4.6$ &
    $5.0$ & $4.9$ \\

    ($\star$) + SkipRNN \cite{campos2018skip}  & $0.09$ & $0.01$ &
    $4.4$ & $4.3$ &
    $5.0$ & $4.6$ \\
    
    ($\star$) + TinyLSTM \cite{fedorov_tinylstm_2020}  & $0.12$ & $0.09$ &
    $4.3$ & $4.4$ &
    $5.0$ & $4.7$ \\

    \bottomrule
  \end{tabular}
\end{table}


\begin{table}[t]
  \caption{Temporal sparsification in the MN-DNN of ($\star$): performance and per-frame cost for updated (w/o skip) and non-updated (w/skip) frames.}
  \label{tab:skip_multi}
  \centering
  \small
  \fontsize{8.5}{9.5}\selectfont
  
  \setlength{\tabcolsep}{4pt}


  \begin{tabular}{lcccccc}
    \toprule

    \multicolumn{1}{c}{} &
    \multicolumn{2}{c}{\textbf{MMACs / frame}} &
    \multicolumn{2}{c}{\textbf{SI-SDR}} &
    \multicolumn{2}{c}{\textbf{SI-SAR}} \\
    \cmidrule(lr){2-3}\cmidrule(lr){4-5}\cmidrule(lr){6-7}

    \textbf{Model} &
    \textbf{w/o Skip} & \textbf{w/ Skip} &
    \textbf{L} & \textbf{R} &
    \textbf{L} & \textbf{R} \\
    \midrule

    MN-DNN of ($\star$) & $0.18$ & $-$ &
    $4.5$ & $4.8$ &
    $5.1$ & $5.0$ \\
    \midrule
    
    ($\star$) + FRS 1/2 & $0.18$ & $0$ &
    $4.4$ & $4.8$ &
    $5.1$ & $5.1$ \\
    
    ($\star$) + FRS 1/4 & $0.18$ & $0$ &
    $4.3$ & $4.8$ &
    $4.8$ & $5.0$ \\
    
    ($\star$) + SkipRNN \cite{campos2018skip} & $0.18$ & $0.02$ &
    $3.8$ & $4.7$ &
    $4.5$ & $5.0$ \\
    
    ($\star$) + TinyLSTM \cite{fedorov_tinylstm_2020}  & $0.21$ & $0.07$ &
    $3.3$ & $4.6$ &
    $3.9$ & $4.8$ \\
    

    \bottomrule
  \end{tabular}
\end{table}

%% file: table4.tex

\begin{table}[t]
  \caption{Effect of window length and type on enhancement performance, evaluated using RT-Tango.}
  \label{tab:lr_metrics_GRNN}
  \centering
  \small
  \fontsize{8.5}{9.5}\selectfont
  


  \begin{tabular}{lccccccc}
    \toprule

    \multicolumn{1}{c}{} &
    \multicolumn{2}{c}{} &
    \multicolumn{2}{c}{\textbf{SI-SDR}} &
    \multicolumn{2}{c}{\textbf{SI-SAR}} \\
    \cmidrule(lr){4-5}\cmidrule(lr){6-7}

    \textbf{aW} &
    \textbf{sW} &
    \textbf{Window} &
    \textbf{L} & \textbf{R} &
    \textbf{L} & \textbf{R} \\
    \midrule

    $32$ & $32$ & sqrtHann &
    $4.4$ & $4.7$ & $5.0$ & $5.0$ \\
    \midrule
    
    $32$ & $8$ & sqrtHann &
    $1.2$ & $1.8$ & $1.4$  & $1.9$\\

    $32$ & $8$ & asyHann &
    $3.7$  & $4.3$  & $4.3$ & $4.5$ \\

    $32$ & $4$ & asyHann &
    $3.4$ & $4.0$ & $3.9$  & $4.1$  \\
    
    \bottomrule
  \end{tabular}
\end{table}

%% file: 5conclusions.tex
\section{Conclusions}
This paper introduced RT-Tango, a distributed binaural SE framework designed to satisfy the stringent real-time, latency, and computational constraints of modern hearing devices. By revisiting the Tango architecture through complementary architectural and signal-processing optimizations, RT-Tango significantly reduces computational complexity while preserving enhancement performance and interaural balance. In addition, RT-Tango-OS, its strictly causal streaming variant, operates with an algorithmic latency of 8 ms, makes RT-Tango particularly well suited for practical deployment in modern binaural hearing aids.

\section{Generative AI Use Disclosure}
The authors used AI tools solely for language editing and clarity improvement. All scientific ideas, content, and results were developed and verified by the authors.